\documentclass[aps,prl,reprint,twocolumn,notitlepage,superscriptaddress,showpacs]{revtex4-1}
\usepackage{graphicx}
\usepackage[table]{xcolor}
\usepackage{amssymb,amsthm,amsmath}
\usepackage{braket}
\usepackage{siunitx}
\usepackage[varg]{txfonts}
\usepackage{upgreek}
\usepackage{dcolumn}
\usepackage{dsfont}
\usepackage{tabularx,colordvi}

\begin{document}
\title{Topological phase transition from trigonal warping in van der Waals multilayers}
\author{Junjie Zeng}
\affiliation{ICQD, Hefei National Laboratory for Physical Sciences at Microscale, and Synergetic Innovation Centre of Quantum Information and Quantum Physics, University of Science and Technology of China, Hefei, Anhui 230026, China}
\affiliation{CAS Key Laboratory of Strongly-Coupled Quantum Matter Physics and Department of Physics, University of Science and Technology of China, Hefei, Anhui 230026, China.}
\author{Yafei Ren}
\affiliation{ICQD, Hefei National Laboratory for Physical Sciences at Microscale, and Synergetic Innovation Centre of Quantum Information and Quantum Physics, University of Science and Technology of China, Hefei, Anhui 230026, China}
\affiliation{CAS Key Laboratory of Strongly-Coupled Quantum Matter Physics and Department of Physics, University of Science and Technology of China, Hefei, Anhui 230026, China.}
\author{Kunhua Zhang}
\affiliation{ICQD, Hefei National Laboratory for Physical Sciences at Microscale, and Synergetic Innovation Centre of Quantum Information and Quantum Physics, University of Science and Technology of China, Hefei, Anhui 230026, China}
\affiliation{CAS Key Laboratory of Strongly-Coupled Quantum Matter Physics and Department of Physics, University of Science and Technology of China, Hefei, Anhui 230026, China.}
\author{Zhenhua Qiao}
\email[Correspondence author: ]{qiao@ustc.edu.cn}
\affiliation{ICQD, Hefei National Laboratory for Physical Sciences at Microscale, and Synergetic Innovation Centre of Quantum Information and Quantum Physics, University of Science and Technology of China, Hefei, Anhui 230026, China}
\affiliation{CAS Key Laboratory of Strongly-Coupled Quantum Matter Physics and Department of Physics, University of Science and Technology of China, Hefei, Anhui 230026, China.}
\date{\today}

\begin{abstract}
  In van der Waals multilayers of triangular lattice, trigonal warping occurs universally due to the interlayer hopping. We theoretically investigate the effect of trigonal warping upon distinctive topological phases, like the quantum anomalous Hall effect (QAHE) and the quantum valley Hall effect (QVHE). Taking Bernal-stacked bilayer graphene as an example, we find that the trigonal warping plays a crucial role in the formation of QAHE in large exchange field and/or interlayer potential difference by inducing extra band inversion points at momentum further away from high symmetric point. The presence of trigonal warping shrinks the phase space of QAHE and QVHE, leading to the emergence of valley-polarized QAHE with high Chern numbers ranging from $ \mathcal{C}=-7 $ to $ 7 $. These results suggest that the universal trigonal warping may play important role when the Bloch states at momentum away from high-symmetric points are involved.
\end{abstract}
	

\maketitle

\textit{Introduction---.} The layered van der Waals materials and their heterostructures have attracted much attention recently, most of which exhibit threefold rotation symmetry~\cite{rev_vdW_13, rev_vdW_16} leading to a triangular distortion of Fermi surface, i.e., the isoenergy lines show threefold rotational symmetry, which is known as trigonal warping~\cite{Akimoto2004, Kechedzhi2007, Ezawa2012, McCann2013}. Such trigonal warping finds itself almost universal existence in the two-dimensional van der Waals layered materials, and brings about various effects of interest~\cite{Dora2009,Rakyta2010,Ortix2012,Mucha2013,Kormanyos2013,Liu2015}. For instance, by theoretically considering the trigonal warping terms, the presence of Majorana fermions in monolayer graphene can be explained in terms of Dirac equation~\cite{Dora2009}. Besides, the Rashba spin-orbit coupling (SOC) induced trigonal warping in monolayer graphene can break up the Fermi circle at low energies as well, one of whose manifestations lies in the trigonal increase in the minimal conductivity~\cite{Rakyta2010}.

Different from monolayer systems where the trigonal warping arises out of differing rationales, the main origins of trigonal warping in multilayer van der Waals materials share a common source, i.e., the interlayer hopping, as reported in various systems, such as in bilayer graphene~\cite{Kechedzhi2007}, bilayer silicene~\cite{Ezawa2012}, and heterostructure of graphene and hexagonal boron nitride~\cite{Ortix2012}. Such trigonal warping arises due to the direct wavefunction overlap between different layers and thus usually leads to a trigonal warping of orders stronger than, for example, that induced by Rashba SOC~\cite{Kuzmenko2009, Rakyta2010, McCann2013} and plays more important role in the electronic structure~\cite{Ortix2012} and transport properties~\cite{Kechedzhi2007}. For example, in bilayer graphene, it takes the responsibility for the unusual behavior of interference effects in electronic transport and a suppression of weak localization effect in the absence of intervalley scattering~\cite{Kechedzhi2007}.

\begin{figure}[htp!]
	\centering
	\includegraphics[width=.48\textwidth]{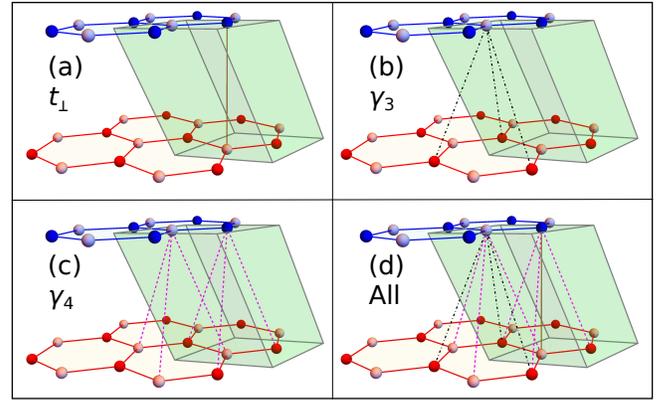}
	\caption{\label{fig:interh_biG}Schematic diagram for interlayer hopping in Bernal-stacked bilayer graphene with the light-green-face parallelepiped the one of primitive cell options and the darkly- and lightly-colored spheres the sublattice $ A $ and $ B $, respectively. With the in-plane hopping terms understood, in Fig.~(a) the perpendicular brown solid line $ t_\perp $ connects between the dimer-site atoms; in Fig.~(b) the black dash-dotted lines establish inter-sublattice hopping for $ \gamma_3 $; in Fig.~(c) the magenta dashed lines build up intra-sublattice hopping characterized by $ \gamma_4 $; and in Fig.~(d) all of the relevant interlayer hoppings are presented.}
\end{figure}

Although the interlayer hopping induced trigonal warping is universal in van der Waals systems of honeycomb lattice, it has not attracted sufficient attention so far in the discipline. In this Rapid Communication, we explore the effect of trigonal warping on the abundant topological phases~\cite{Ren2016} in Bernal-stacked bilayer graphene, e.g., QAHE~\cite{Ding2011, Qiao2012, Jiang2012, Chang2013, Pan2014, Qiao2014, Bestwick2015, Ren2016a} and QVHE~\cite{Ding2011, Xiao2007, Jung2011, Qiao2011_Highway, Qiao2013}. We find that, by influencing the band structure away from high symmetry points in the higher energy regime, the trigonal warping not only changes the Chern number of some topological phases, but also enriches topological phases inducing QAHE of high Chern number $\mathcal{C}=\pm7 $. These results suggest that the influence of trigonal warping on the electronic structure as well as transport properties shall to be considered when the Fermi surface involves the momentum away from high-symmetric points.

\begin{figure*}[htp!]
	\centering
	\includegraphics[width=1.\textwidth]{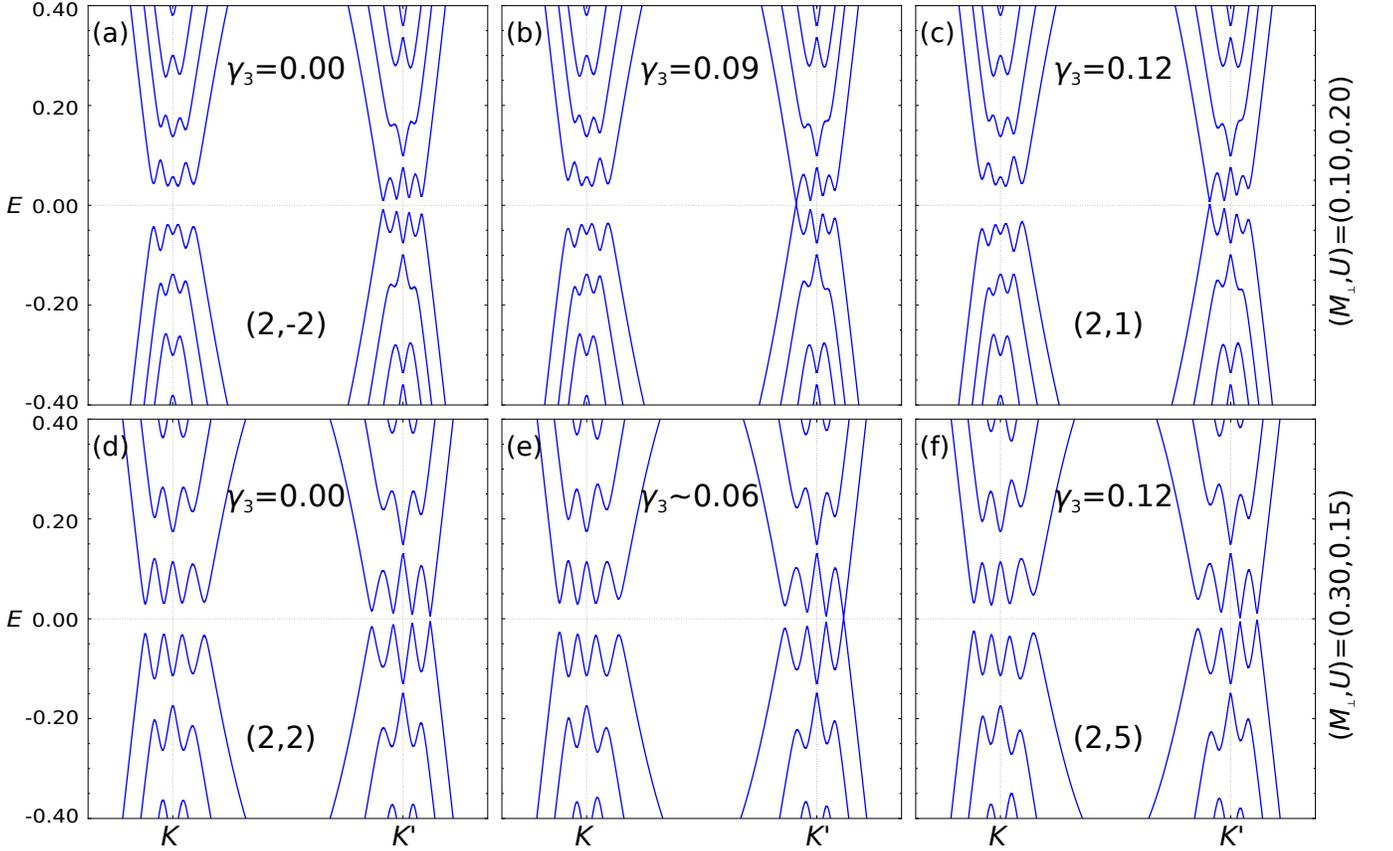}
	\caption{\label{fig:pha_trans}Band structure evolution induced by increasing trigonal warping from zero in gated bilayer graphene (the two numbers in parentheses are Chern numbers contributed from $ K $ and $ K' $ valleys, i.e., $ (\mathcal{C}_K,\mathcal{C}_{K'}) $). Here,  we choose two groups of $ (M_\perp,U) $ and the strength of extrinsic-Rashba SOC is $ t_\text{SO}=0.04 $. Figs.~(a)-(c) show the process of topological phase transition induced by trigonal warping located at $ (M_\perp,U)=(0.10,0.20) $ in the parameter space with a QVHE phase as the initial phase. Analogously, Figs.~(d)-(f) depict the phase transition start from a QAHE phase with $ (M_\perp,U)=(0.30,0.15) $. Notice that in Fig.~(b) the bands close on $ M $-$ K' $ when the trigonal warping strength $ \gamma_3=0.09 $, whereas in Fig.~(e) the Dirac cone is formed on $ K' $-$ \mathit\Gamma $ when $ \gamma_3=0.056 $. And it is always true that $ \gamma_4=0.37\gamma_3 $, see the rationality for this fixed ratio in the main text.}
\end{figure*}

\textit{System Model Hamiltonian---.} Our numerical calculation is based on the tight-binding model Hamiltonian of Bernal-stacked bilayer graphene in the presence of uniform exchange field and Rashba SOC. To capture the physics of trigonal warping, the next-nearest-neighbor interlayer hoppings are included and the corresponding Hamiltonian can be expressed as~\cite{McCann2013}:
\begin{align}
H_\text{BLG}=&\sum_{l=\text{B,T}}H_\text{SLG}^l+t_\perp\sum_\alpha\sum_{\substack{\langle i\in\text{B}(B)\\j\in\text{T}(A)\rangle}}(c_{i\alpha}^\dagger c_{j\alpha}+\text{H.c.})\nonumber\\
&+\left(\gamma_4\sum_{\substack{\alpha,\\\varsigma=A,B}}\sum_{\substack{\langle\langle i\in\text{B}(\varsigma)\\j\in\text{T}(\varsigma)\rangle\rangle}}-\gamma_3\sum_\alpha\sum_{\substack{\langle\langle i\in\text{B}(A)\\j\in\text{T}(B)\rangle\rangle}}\right)(c_{i\alpha}^\dagger c_{j\alpha}+\text{H.c.})\nonumber\\
&+U\sum_{\substack{\alpha,i\in\text{B}\\j\in\text{T}}}(c_{i\alpha}^\dagger c_{i\alpha}-c_{j\alpha}^\dagger c_{j\alpha}),\label{eq:Hami}
\end{align}
where the first term $ H_{\text{SLG}}^{\text{B,T}}$ is the nearest-neighbor tight-binding Hamiltonian of the bottom/top graphene layer in the presence of the extrinsic-Rashba SOC and the exchange field with strengths of $ t_\text{SO} $ and $ M_\perp $ respectively, which is expressed as~\cite{Qiao2010,Qiao2011}:
\begin{align}
H_\text{SLG}=&-t\sum_{\braket{ij}\alpha}c_{i\alpha}^\dagger c_{j\alpha}+\mathrm{i} t_\text{SO}\sum_{\substack{\braket{ij}\\\alpha\beta}}c_{i\alpha}^\dagger({\boldsymbol{s}}\times{\hat{\boldsymbol{d}}_{ij}})_{\alpha\beta}^z c_{j\beta}\nonumber\\
&+M_\perp\sum_{i\alpha}c_{i\alpha}^\dagger s_z c_{i\alpha}.
\end{align}
The second term of Eq.~\eqref{eq:Hami} represents the dimer coupling between $ B $ sites of the bottom layer and the corresponding $ A $ sites of the top layer, as displayed in Fig.~\ref{fig:interh_biG}(a). The trigonal warping terms in bilayer graphene originates from interlayer next-nearest-neighbor hopping as portrayed by the third term, with $ \gamma_{3,4} $ corresponding to the hopping strengths between different kinds of atomic sites as displayed separately in Figs.~\ref{fig:interh_biG}(b) and \ref{fig:interh_biG}(c). The last term describes the interlayer potential energy difference $ U $. Hereafter, we take the nearest-neighbor hopping energy $ t $ as the energy unity. The following parameters are set as $ t_\perp=0.12, \gamma_4=0.37\gamma_3 $ (a ratio between their practical values $ \gamma_3=\SI{.38}{\electronvolt} $ and $ \gamma_4=\SI{.14}{\electronvolt} $), and the extrinsic-Rashba SOC $ t_\text{SO}=t_\perp/3 $ unless otherwise noted~\cite{Kuzmenko2009,McCann2013}.

\begin{figure*}[htp!]
	\centering
	\includegraphics[width=\textwidth]{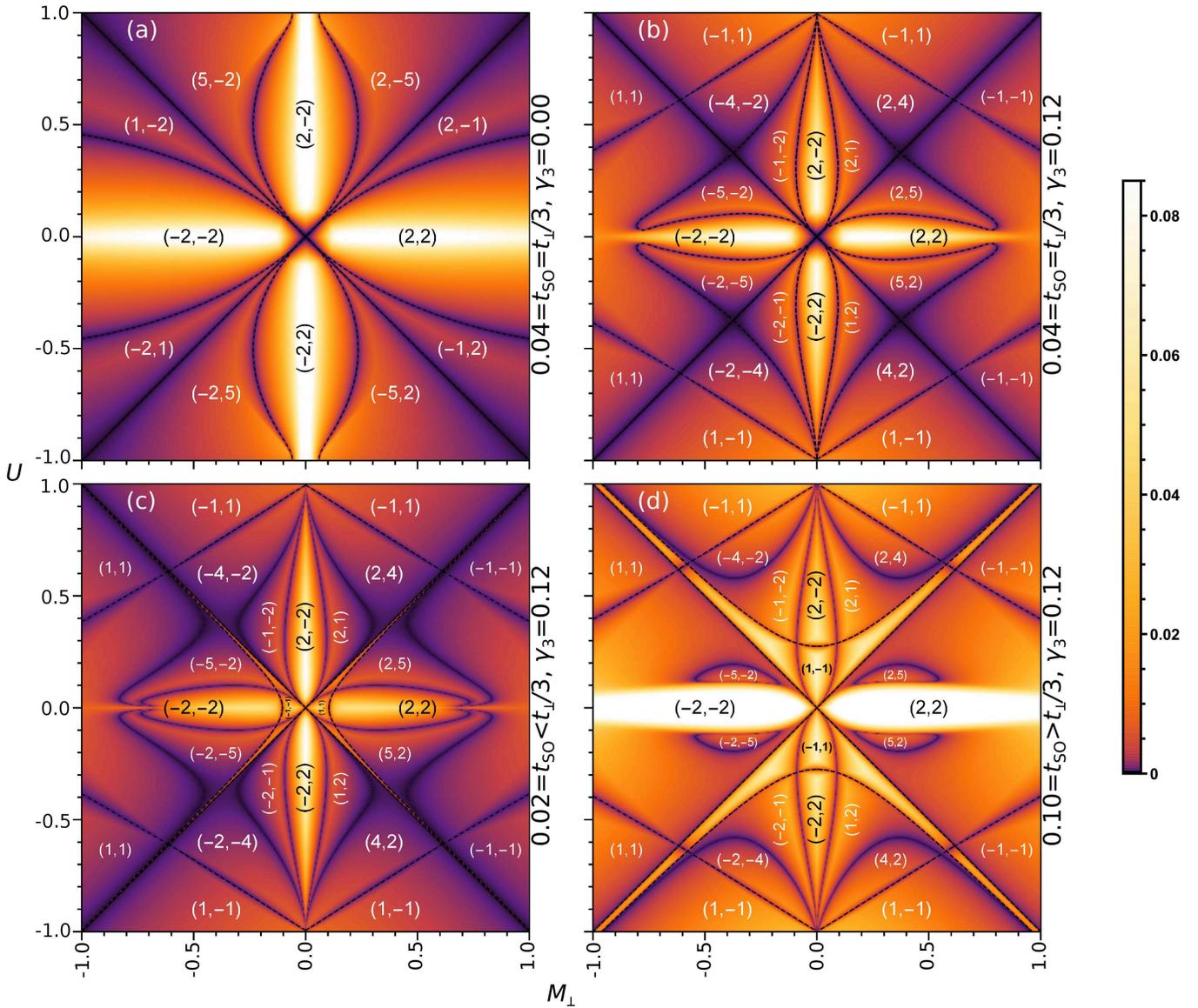}
	\caption{\label{fig:pha_diag} Comparison of phase diagrams on $ M_\perp $-$ U $ plane of bilayer graphene. The color bar indicates the magnitude of band gap, and the two integers in parentheses represent the Chern numbers contributed from valleys $ K $ and $ K' $ $ (\mathcal{C}_K,\mathcal{C}_{K'}) $, respectively. Fig.~(a) displays the simplest case, wherein even though without trigonal warping ($ \gamma_3=0.00, t_\text{SO}=0.04 $) there exist two valley-polarized phases other than the QVHE and QAHE phases. Fig.~(b) shows the effect of trigonal warping ($ \gamma_3=0.12, t_\text{SO}=0.04 $) based on Fig.~(a) that each adiabatically disconnected regions are reshaped and new phases emerge, among which are those with a high Chern number $ \pm 7 $. Figs.~(c) and (d) show that when $ t_\text{SO} $ is deviated from $ 0.04 $, the hyperbola phase borders appear, however, with the main features depicted in Fig.~(b) maintained.}
\end{figure*}

\textit{Band structure analysis---.} Through Fourier transform, we can express Eq.~\eqref{eq:Hami} in the momentum space spanned by the basis of $ \set{\ket{(\boldsymbol{k})l\varsigma\alpha}} $ with $ l\in\set{\text{B,T}}, \varsigma\in\set{A,B}\text{ and }\alpha\in\set{\uparrow,\downarrow} $ labeling the bottom/top layer, the $ A/B $ sublattice and the spin up/down states, respectively. We obtain the band structure by directly diagonalizing this $ 8\times 8 $ Hamiltonian. The band structure of bilayer graphene shows gapless quadratic band crossings at $ K/K' $ valleys and the introduction of either interlayer potential difference or the exchange field and the extrinsic-Rashba SOC can open up a band gap, which holds either QVHE or QAHE phase. The competition between these two effects gives rise to rich topological phases as reported in Refs.~\cite{Tse2011,Qiao2013}.

Now we demonstrate the effect of the trigonal warping terms on these two topological phases by calculating the band structure as shown in Fig.~\ref{fig:pha_trans}. In the absence of the trigonal warping term, the system exhibits a QVHE phase when $ |M_\perp|<|U|$ as shown in Fig.~\ref{fig:pha_trans}(a), where the Chern numbers of $ K/K' $ valleys are respectively $ \pm 2 $. Although it leaves the band gap of $ K $ valley nearly constant, the increase of the strength of trigonal warping changes that of $ K' $ valley, which first reduces to zero and then reopens as plotted in Figs.~\ref{fig:pha_trans}(b) and \ref{fig:pha_trans}(c) with the critical value being $ \gamma_3=0.09 $ that is smaller than the physical value \cite{McCann2013}. Such a band closing and reopening indicates a topological phase transition that changes the Chern number of $ K' $ valley by three. Therefore, the system exhibits a valley-polarized QAHE phase with a Chern number of $ \mathcal{C}=\mathcal{C}_K+\mathcal{C}_{K'}=3 $ and a valley Chern number of $ \mathcal{C}_{\rm{V}}=\mathcal{C}_K-\mathcal{C}_{K'}=1 $. It is noteworthy that, when the topological phase of this system changes from QVHE to QAHE, the band gap closing occurs at momentum away from $K/K'$ points, which is different from the phase transition at low-energy regime by changing $U$ or $M_\perp$ where the band gap closes at $K/K'$ point when $|U|=|M_\perp|$ as reported in Ref.~\onlinecite{Qiao2013}. This indicates that the trigonal warping plays crucial role at momentum away from high-symmetric point.

Similarly, for the QAHE phase when $ |M_\perp|>|U|$, we also find that the presence of the TW terms can induce a topological phase transition at $K'$ valley, which changes the Chern number from $ (\mathcal{C}_K,\mathcal{C}_{K'})=(2,2) $ to $ (2,5) $ leading again to a valley-polarized QAHE phase with a high Chern number up to 7 as shown in Figs.~\ref{fig:pha_trans}(d)-\ref{fig:pha_trans}(f). Furthermore, for both cases, the increment in the Chern numbers is three due to the $ \text{C}_3 $ symmetry of the system, which makes the band closing occur at three different points in the Brillouin zone simultaneously at the critical point. Then it fairly can be seen that each single band crossing contributes to a unit Chern number variation. However, the band closing points for QVHE and QAHE locate at different high-symmetric lines of $ M $-$ K' $ and $ K' $-$ \mathit\Gamma $, respectively [see Figs.~\ref{fig:pha_trans}(b) and \ref{fig:pha_trans}(e)]. In addition, one also notes that the critical trigonal warping strengths for the topological phase transitions differ for different parameters, indicating that the topological phases depend on the specific values of the trigonal warping strength. It is worth restating that, in the above calculations, the proportionality between $ \gamma_3 $ and $ \gamma_4 $ is fixed. This choice is validated by the further investigation into the case of changing $ \gamma_4 $ independent of $ \gamma_3 $, in which we find that it can only shift the low-energy region of the band structure slightly without inducing new topological phase.

\textit{Phase diagram---.} To fully illustrate the phase transitions induced by the trigonal warping, we calculate the phase diagrams with and without trigonal warping for different extrinsic-Rashba SOC strengths of $ t_\text{SO}=0.02 $, $ 0.04 $, and $ 0.10 $. Without loss of generality, we first study the phase diagrams of $ t_\text{SO}=0.04 $ without and with trigonal warping as shown in Figs.~\ref{fig:pha_diag}(a) and \ref{fig:pha_diag}(b), respectively. In the absence of trigonal warping, one can find that, in the regions with small $ M_\perp $ and $ U $, the QVHE with valley Chern number $ \mathcal{C}_\text{V}=\pm4 $ and vanishing Chern number $ \mathcal{C} $ occurs with when $ |M_\perp|<|U| $, whereas the QAHE phase with $ \mathcal{C}=\pm4 $ appears when $ |M_\perp|>|U| $, which reveals the phase diagrams reported in previous work by using low-energy effective Hamiltonian~\cite{Qiao2013,Tse2011}. However, the situation becomes quite different for larger $ M_\perp $ and $ U $. First, the valley-polarized QAHE appears with Chern numbers $ (2,-5) $ and $ (2,-1) $ even without the trigonal warping as Fig.~\ref{fig:pha_diag}(a) depicts. It is noteworthy that the sign change of $ U $ leads to an interchange of the Chern numbers at $ K $ and $ K' $ valleys: $ \mathcal{C}_K\leftrightarrow\mathcal{C}_{K'} $, while the sign change of $ M_\perp $ induces a sign flip of Chern number: $ \mathcal{C}\rightarrow-\mathcal{C} $. These rules are also valid in the presence of trigonal warping.

The phase diagram in the presence of trigonal warping is presented in Fig.~\ref{fig:pha_diag}(b). By comparing Figs.~\ref{fig:pha_diag}(a) and \ref{fig:pha_diag}(b), one can find that the phase diagram has been strongly influenced. Specifically, the phase spaces for the QVHE with $ (\mathcal{C}_K,\mathcal{C}_{K'})=(2,-2) $ and the QAHE with $ (\mathcal{C}_K,\mathcal{C}_{K'})=(2,2) $ are effectively diminished while various new topological phases arise. In the regions near $ |M_{\perp}|=| U| $, the topological phase transition changes the Chern number by six. For instance, the QAHE of $ (\mathcal{C}_K,\mathcal{C}_{K'})=(2,-5) $ and $ (2, -1) $ become changed to be $ (2, 1) $ and $ (2, 5) $, respectively. In the parameter space away from the quadrant bisectors, however, the Chern number is changed by three. Moreover, if $ M_\perp $ and $ U $ are so large that are comparable to the in-plane hopping energy, another three new topological phases with $ (\mathcal{C}_K,\mathcal{C}_{K'})=\pm(2, 4) $ and $ \pm(-1,-1) $, as well as the QVHE with $ (\mathcal{C}_K,\mathcal{C}_{K'})=\pm(-1, 1) $ occur.

In previous calculations, we set the extrinsic Rashba SOC strength $ t_\text{SO}=0.04 $ , which equals to $ t_{\perp}/3 $ for simplicity~\cite{Qiao2013}. When $ t_{\rm{SO}} $ deviates from this value, new topological phases arise. However, this does not influence the major characters of the phase diagram. As displayed in Figs.~\ref{fig:pha_diag}(c) and \ref{fig:pha_diag}(d), we plot the phase diagram for $ t_{\rm{SO}}<t_{\perp}/3 $ and $ t_{\rm{SO}}>t_{\perp}/3 $, respectively. The decrease (increase) of $ t_{\rm{SO}} $ introduces new topological phases in QAHE (QVHE) regime near the center of phase diagram, which is in consistent with the previous work in Ref.~\onlinecite{Qiao2013}. Moreover, near the cross-like phase border of $ (M_{\perp}, U)\approx(\pm 0.38, \pm 0.38) $, the variation of $ t_{\rm{SO}} $ also splits the crossing phase border horizontally and vertically, respectively. Nevertheless, there is not new topological phase arising.

\textit{Summary and Discussion---.} In conclusion, we study the effect of trigonal warping on the topological nontrivial phases in Bernal stacked bilayer graphene, which is induced by the next-nearest-neighbor interlayer hopping. We find that the trigonal warping plays a crucial role in determining the topological phases of bilayer graphene induced by exchange field $ M_{\perp} $ and the interlayer potential difference $ U $. Specifically, the presence of trigonal warping can modify the QAHE and QVHE phase by inducing topological phase transition that changes the Chern number by three or six. For instance, the QAHE phase in the absence of trigonal warping with a Chern number of $ (\mathcal{C}_K,\mathcal{C}_{K'})=(2,2) $ is changed by including the trigonal warping to a new phase with Chern number $ (2,5) $, corresponding to a valley-polarized QAHE phase exhibiting a rather high Chern number. What's more, because the origin of trigonal warping considered is the interlayer hopping without any other conditions specified, a generalization can be drawn that in other multilayer van der Waals systems, the interlayer hopping induced trigonal warping may also drive new phenomena.

\textit{Acknowledgements---.} This work was financially supported by NNSFC (11474265), the China Government Youth 1000-Plan Talent Program, Fundamental Research Funds for the Central Universities (WK3510000001 and WK2030020027), and the National Key R \& D Program (2016YFA0301700). The Supercomputing Center of USTC is gratefully acknowledged for the high-performance computing assistance.

	\bibliographystyle{apsrev4-1}
	
\end{document}